\documentstyle[12pt,epsf,epsfig,rotate]{article}

\newcommand {\blackboardrrm}{\mathchoice
{\rm I\kern-0.21 em{R}}{\rm I\kern-0.21 em{R}}
{\rm I\kern-0.19 em{R}}{\rm I\kern-0.19 em{R}}}

\newcommand {\blackboardzrm}{\mathchoice
{\rm Z\kern-0.32 em{Z}}{\rm Z\kern-0.32 em{Z}}
{\rm Z\kern-0.28 em{Z}}{\rm Z\kern-0.28 em{Z}}}

\newcommand {\si} {\sigma}
\newcommand {\be}  {\begin{equation}}
\newcommand {\ee}  {\end{equation}}
\newcommand {\bea} {\begin{eqnarray} \nonumber }
\newcommand {\eea} {\end{eqnarray}}
\newcommand {\lan} {\langle}
\newcommand {\ran} {\rangle}

\renewcommand {\d} {{\mbox d}}

\newcommand{\overl}[1]{\overline{#1}}

\begin{document}

\title{On the Phase Structure of the $3D$ Edwards Anderson Spin Glass}

\author{Enzo Marinari$^{(a)}$, Giorgio Parisi$^{(b)}$
 and Juan J. Ruiz-Lorenzo$^{(b)}$\\[0.5em]
$^{(a)}$  {\small  Dipartimento di Fisica and INFN, 
Universit\`a di Cagliari}\\
{\small   \ \  Via Ospedale 72, 07100 Cagliari (Italy)}\\[0.3em]
{\small   \tt marinari@ca.infn.it}\\[0.5em]
$^{(b)}$  {\small  Dipartimento di Fisica and INFN, Universit\`a di Roma}
   {\small {\em La Sapienza} }\\
{\small   \ \  P. A. Moro 2, 00185 Roma (Italy)}\\[0.3em]
{\small   \tt giorgio.parisi@roma1.infn.it  ruiz@chimera.roma1.infn.it}\\[0.5em]
}

\date{February 17, 1998}

\maketitle

\begin{abstract}

We characterize numerically the properties of the phase transition of
the three dimensional Ising spin glass with Gaussian couplings and of
the low temperature phase.  We compute critical exponents on large
lattices.  We study in detail the overlap probability distribution and
the equilibrium overlap-overlap correlation functions.  We find a
clear agreement with off-equilibrium results from previous work.
These results strongly support the existence of a continuous
spontaneous replica symmetry breaking in three dimensional spin
glasses.
\end{abstract}

\vfill
PACS mumbers: 75.50.Lk, 05.50.+q, 64.60.Cn

\thispagestyle{empty}
\newpage

\section{\protect\label{S_INT}Introduction}

Recent numerical simulations \cite{KAWYOU,BERJAN} have given a strong 
numerical evidence of the existence of a spin glass phase transition 
in the $3D$ Edwards Anderson spin glass (for a critical point of view 
see \cite{NS,OUR,WINDOW}). 
The first studies of such models are today $15$ years 
old (for a recent review see \cite{BOOK}), and the Replica Symmetry 
Breaking (RSB) mean field solution \cite{MEPAVI} is found to describe 
the most of the properties observed in finite dimensional models.  The 
first issue is the study of the probability distribution of the order 
parameter $P(q)$, but the innovative features of the RSB solution 
inspire many questions that can be answered numerically.

Recently many of these issues have indeed been analyzed numerically.  
Correlation functions and block overlaps of the $3D$ spin glass have 
been shown in \cite{MAPARURI} to have a mean field like behavior, the 
$4D$ model has been studied in detail in \cite{4DIM}, and the role of 
the upper critical dimension, $D_{c}^{U}=6$, has been analyzed in 
\cite{MEANFIELD}. It is also remarkable that a deep relation among 
static and dynamic behavior has be shown to be valid, for the $3D$ 
and $4D$ models, in \cite{DINAMI}. At last it is worth to remind the 
reader that results about the transition in field in finite 
dimensional models have been obtained in \cite{FIEUNO,FIEDUE}.

Here we try to push these results to the limit allowed by the 
computers we can use today, and by the numerical improved algorithms 
we use for our simulations.  We consider the $3D$ spin glass with 
gaussian couplings, that make safer the approach to the low $T$ region 
(because of the absence of accidental degeneracy) and allow to check 
for universality when comparing to the results of 
\cite{KAWYOU,BERJAN}.  After defining our model and our statistical 
quantities and giving details about the numerical simulations we 
determine critical exponents by analyzing the overlap Binder cumulant 
and the spin glass susceptibility in section (\ref{S_BIN}).  We 
analyze in detail the behavior of $P(q)$ in (\ref{S_PQ}), by checking 
in detail the presence of a mean field like behavior.  We analyze the 
behavior of a simple function in (\ref{S_SUM}).  In section 
(\ref{S_COR}) we are able to compute equilibrium correlation function, 
for all states and for the zero overlap sector of the theory, and to 
show that these results coincide with the one obtained by using an 
extrapolation in \cite{MAPARURI}.

\section{\protect\label{S_MOD}The Model}

We consider a three dimensional ($3D$) Edwards Anderson spin glass 
model with Gaussian quenched random couplings $J$.  The model is 
defined on a simple cubic lattice with periodic boundary conditions.  
The Hamiltonian of the system is

\be
  {\cal H} \equiv -\sum_{<ij>} \sigma_i J_{ij} \sigma_j\ ,
\ee
where by $<ij>$ we indicate that the sum runs
over couples of first neighboring sites. The $J_{ij}$ are 
quenched Gaussian variables with zero mean and unit variance.
We consider two real replicas of the system (in the same realization 
of the disorder). We define the overlap among the two real replicas 
$\alpha$ and $\gamma$ at site $i$ as

\be
  q_{J,i}^{\alpha,\gamma}
  \equiv \sigma_i^{\alpha}   \sigma_i^{\gamma}\ ,
  \protect\label{E-QI}
\ee
(where the subscript $J$ reminds us that we are in a fixed realization 
of the quenched disorder $J$), and a total overlap 

\be
  q^{\alpha,\gamma}_J \equiv \frac{1}{V}\sum_i q_i^{\alpha,\gamma}\ ,
  \protect\label{E-QSUM}
\ee
where we will frequently ignore the superscripts, and denote it
by $q$.  Its probability distribution for a given sample (defining
$\lan \ldots \ran$ the thermodynamical average) is

\be 
  P_J(q) \equiv \lan \delta(q-q^{\alpha,\gamma}_J )\ran\ ,
  \protect\label{E-PQJ}
\ee 
and averaging over samples (denoting with
$\overl{\cdot\cdot \cdot}$ the average on the disorder) we define

\be 
  P(q) \equiv \overl{P_J(q)}\ .  
  \protect\label{E-PQ}
\ee 
The overlap Binder parameter is defined as

\be
  g \equiv \frac12 
  \left[
  3-\frac{\overline{\langle q^4\rangle}}{\overline{\langle 
  q^2\rangle}^2} \right] \ .
  \protect\label{E-BINDER}
\ee
The spatial overlap-overlap correlation function is

\be
  C_{i,j} \equiv \overline{
  \langle q_i q_{i+j} \rangle}
  =  \overline{
  \langle \sigma_i \tau_i \sigma_{i+j} \tau_{i+j}\rangle}
  =  \overline{
  \langle \sigma_i \sigma_{i+j}\rangle^2}\ .
  \protect\label{E-GIJ}
\ee
We will denote by $C_j$ (or $C(j)$) the $q-q$ correlation at distance 
$j$, averaged over different site couples whose distance is $j$.

It is also possible and interesting to define a correlation function 
where the overlap is fixed to a given value: one selects couples of 
equilibrium configurations (the two real replicas that constitute our 
system) with a given fixed overlap $q$ and compute the 
correlation function among these configurations.  We will 
denote these correlation functions with the symbol $C_q(x)$.  
Obviously

\be
C(x)=\int dq\  P(q)\  C_q(x)\ ,
\ee
i.e. $C(x)$ is the sum of $C_q(x)$ weighted with the static
probability distribution of the overlaps. Finally we will define
the connected correlation functions, by using the symbols ${\hat C}_q(x)$
and ${\hat C}$,

\bea
  {\hat C}_q(x) &=& C_q (x)- q^2 \ , \\
  {\hat C}(x)   &=& C(x)-\overline{\lan q^{2} \ran} \ .
\eea
Some crucial properties of the propagators, as computed in the RSB 
framework, will be useful in the following (see \cite{DEKOTE} and 
references therein).  At tree level one finds that (in dimension 
$D<D^{u}_{c}=6$ there are corrections, and the contributions due to 
the anomalous dimension has to be included)

\begin{equation}\label{correlations}
{\hat C}_q(x)\propto\left\{
\begin{tabular}{c c c}
$x^{4-D}$& if &  $q=0$\ ,\\
$x^{3-D}$& if &  $0<q<q_{\rm EA}$\ ,\\
$x^{2-D}$& if &  $q=q_{\rm EA}$\ .\\
\end{tabular}
\right.
\end{equation}
We have based the analysis of the numerical results of our numerical 
simulations on finite size scaling techniques.  When changing the 
temperature $T$ and the lattice size $L$ the overlap susceptibility 
$\chi\equiv V \overline{\lan q^{2} \ran}$ scales, in all generality, 
as

\be
  \chi(L,T)=L^{\frac{\gamma}{\nu}}\left(h_1\left(L^{1/\nu} (T-T_c)\right)
  +L^{-\omega} h_2\left( L^{1/\nu} (T-T_c)\right) 
  + O(L^{-2\omega}) \right) \ ,
\ee
and the overlap Binder parameter scales as

\be
g(L,T)=f_1\left(L^{1/\nu} (T-T_c)\right) +L^{-\omega} f_2\left(
L^{1/\nu} (T-T_c)\right) +  
O(L^{-2\omega}) \ ,
\ee
where $\omega$ is the exponent that determines the corrections to the 
scaling (it is the derivative of the $\beta$-function at the critical 
coupling, \cite{GIORGIO}), and $f_1,f_2,h_1$ and $h_2$ are universal 
functions.  In particular this effect is important for the overlap Binder 
parameter at $T=T_{c}$, and gives scaling violations at the infinite 
volume critical point.  One has that

\be
  g(L,T_c)=f_1(0) +L^{-\omega} f_2(0) + O(L^{-2\omega})\ .
\ee
This effect explains why there is not an unique crossing point of the Binder
cumulant curves (see, for instance, reference \cite{PARU}), and has to 
be considered in the analysis of the numerical data.
A one loop field theoretical computation \cite{DeAlcantara} gives

\be
  \omega=6-D \ ,
\ee
i.e.  in $D=3$ at first order in perturbation theory one expects 
$\omega =3$.  Of course this number will be modified by 
renormalization, but we expect the correct result not to be far from 
the previous guess.  In a related theory, the four dimensional site 
percolation, also described by a generalized $\phi^3$ theory, an 
exponent for the corrections to scaling $\omega \simeq 1.13$ has been 
determined numerically.  That has to be compared to the naive (one loop) 
expectation, $\omega=2$, and to the three loop value (with Pad\'e 
resummation), $\omega=1.52$ \cite{PERCO}.  In spin glasses the  
computer power available at present does not allow an accurate 
estimate of the correction to scaling exponent.

\section{\protect\label{S_EQ}Numerical Methods}

We will base our discussion on large scale simulations of three 
dimensional systems of linear size $L=4,6,8,10,12$ and $16$.  The 
(earlier) simulations on smaller lattices use the tempering updating 
scheme \cite{TEMPERING}, while (more recent) simulations on larger 
lattices use the parallel tempering updating scheme \cite{TEMPERING1}, 
which turns out to be very effective (see \cite{ENZO} for a review of 
improved methods, and \cite{BERJAN} for recent work relevant for spin 
glasses). In all cases checking that the systems were, as far as we 
could establish, fully thermalized, has been the first of our worries. 

We have used the APE-100 parallel supercomputer \cite{APE}.  Some of 
the first runs have used the {\em tube} version (with $128$ 
processors), while the bulk of the simulations have been run on the 
{\em tower} version, with $512$ processors, with a peak performance of 
$25$ Gflops.  On the {\em tower} our codes has a sustained speed of 
$6$ Gflops, and updates $200$ million spins per second (even if spins 
can only take two values, the model is based on Gaussian couplings, 
and floating point arithmetics is needed).

The simulated tempering method \cite{TEMPERING,TEMPERING1} (together 
with multicanonical approaches, see \cite{BERJAN} and references 
therein) is a quantum leap for simulations of systems with a complex 
phase space.  One can now thermalize, in the broken phase, systems 
that could never be studied with normal Monte Carlo.  Normal tempering 
works quite easily up to $L=10$ at $T\simeq 0.70 T_c$ (even if the 
fine tuning of the $g$ constants of \cite{TEMPERING} is somehow 
cumbersome).  For larger lattices, $L=12$ and $16$, we have found the 
use of parallel tempering \cite{TEMPERING1} mandatory.  As a negative 
{\em caveat} we feel like adding that using $2$ months of CPU of our 
$25$ Gflops computer we have not been been able to thermalize a $L=24$ 
lattice in the low temperature phase (down to $0.7 T_c$).

The common characteristic of tempering and parallel tempering is that 
the temperature becomes a dynamical variable.  In the simulated 
tempering we propose to update the temperature of the system at run 
time, after one or more usual Monte Carlo sweeps of all the spins.  
This fact makes possible to go from the paramagnetic phase to the spin 
glass phase and back; these changes enable the system to avoid the 
high free energy barriers that separate pure states, making 
possible to escape from the dynamical traps (the metastable states).

The parallel tempering is far easier to program than the simulated 
tempering.  While in the simulated tempering we need to estimate with 
good precision, in a series of thermalization runs, the relative free 
energy of two systems at two contiguous temperatures (in the set of 
the allowed $T$ values), in the parallel tempering method this problem 
is solved by the dynamics itself (i.e.  the method generates, at run 
time, the correct free energies).  For more details see references 
\cite{TEMPERING,TEMPERING1,BOOK,ENZO}.  For sake of completeness we 
give here the scheduling of the two approaches.  In the simulated 
tempering method the process is:

\begin{enumerate}
        
\item We perform $N_{\rm TERM}$ iterations at a fixed temperature $T_{A}$
using the Metropolis algorithm.

\item We perform $N_{\rm FE}$ steps with the Metropolis algorithm at 
the same temperature $T_{A}$ that will be used to compute a first 
guess for the relative free energies.

\item We repeat the steps $1$ and $2$ following an annealing procedure
(i.e. we start with the highest temperature and finish with the
coldest using the final configuration of a given temperature as
the initial configuration for the next lower temperature).

\item We perform five cycles of Metropolis plus simulated tempering 
runs (five series of one Metropolis sweep of all the lattice spins at 
fixed $T$ plus one trial $T$ update), at the end of each we tune the 
values of the relative free energy to make the system visiting each 
allowed $T$ value for the same time.  The total number of steps (for 
the five cycles) is $N_{\rm UFE}$

\item Finally we perform $N_{\rm MEASURES}$ steps, measuring the 
interesting physical quantities.

\end{enumerate}

\begin{table}
\centering
\begin{tabular}{|c||c|c|c|c|c|} \hline
$L$ &  Samples & $N_{\rm TERM}$  & $N_{\rm FE}$  & $N_{\rm UFE}$
& $N_{\rm MEASURES} $ \\ \hline \hline
4   & 33200   & 10000  & 10000  & 50000  & 10000\\\hline
4   & 2048    & 200000 & 350000 & 2000000 & $10^6$ \\\hline
6   & 2048    & 200000 & 500000 & $2.5\times 10^6$ & $10^6$\\\hline
8   & 512     & 100000 & 250000 & 500000   & $10^8$  \\\hline
10  & 512     & 200000 & 350000 & $10^6$   &$10^7$  \\ \hline
\end{tabular}
\caption[0]{Parameters of the simulated tempering runs.
\protect\label{table:sim}}
\end{table}

We report in table (\ref{table:sim}) the values of the parameters
used in the simulated tempering runs. 

As we already said parallel tempering is simpler.  Here there are no 
free parameters to be fine tuned.  We first run $N_{\rm TERM}$ 
thermalization sweeps of the $N_{\beta}$ copies of the system (each at 
a different $T$ value), using from the start a Monte Carlo sweep for 
the spins followed by trial temperature sweeps among the copies.  
After thermalizing we run $N_{\rm MEASURES}$ sweeps, like the ones we 
just described, during which we measure and average the physical 
quantities.  We show the parameters of the parallel tempering in table 
(\ref{table:par}).  For the $L\le 12$ lattice we have used temperatures 
ranging from $1.3$ to $0.7$ with a step of $0.05$, while that for the 
$L=16$ lattice we have taken $T$ ranging from $1.8$ down to $0.7$ with 
a step of $0.05$.

\begin{table}
\centering
\begin{tabular}{|c||c|c|c|} \hline
$L$ &  Samples & $N_{\rm TERM} $  & $N_{\rm MEASURES} $\\ \hline\hline
8   & 4096     & 400000                 & $2 \times 10^6$ \\ \hline
10  & 2048     & 500000                 & $10^6$         \\\hline
12  & 2048     & $10^6$                 & $10^6$         \\\hline
16  & 900      & $10^6$                 & $10^6$         \\\hline
\end{tabular}
\caption[0]{
\protect\label{table:par} Parameters of the parallel tempering runs.}
\end{table}

As we said checking thermalization has been one of the crucial issues 
of this work. We have checked the following facts:

\begin{enumerate}

\item In the both updating schemes the time that a system spends with
a given temperature must be independent of the temperature. By
construction this fact holds only at equilibrium. All our runs verify
this property.

\item The acceptance factor for the temperature update has been 
monitored, and kept in the range $0.2-0.5$.

\item A set of relations among expectation value has been recently 
proven by Guerra \cite{GUERRA} (see also \cite{AIZCON,PARSIX}). These 
relations are satisfied at equilibrium. All our measurements satisfy 
them with a small error.

\item The single sample probability distributions of the order 
parameter $P_{J}(q)$ have to be symmetric for $q\to -q$. We have 
verified this symmetry.

\item We have monitored the growth of the spin glass susceptibility 
and the overlap Binder cumulant.  The times that these quantities need to 
reach their plateau values are estimates of thermalization times.  We 
have kept them under control.

\end{enumerate}

\section{\protect\label{S_BIN}The Binder Cumulant and the Susceptibility}

We will start by discussing our analysis of the overlap Binder 
cumulant, supporting the existence of a spin glass phase transition 
\cite{KAWYOU,BERJAN}.  We show our data in figure (\ref{fig:binder1}).  
A magnified view is in figure (\ref{fig:binder2}): it makes clear the 
crossing (the signature of a phase transition) between the $L=4$ and 
$L=16$ curves and the $L=8$ and $L=16$ curves.  From figure 
(\ref{fig:binder2}) a first (qualitative) guess for the critical 
temperature is

\be 
  T_c \simeq 0.98\pm 0.05 \ .  
\ee 

\begin{figure}[htbp]
\begin{center}
\leavevmode
\epsfysize=250pt
\epsffile{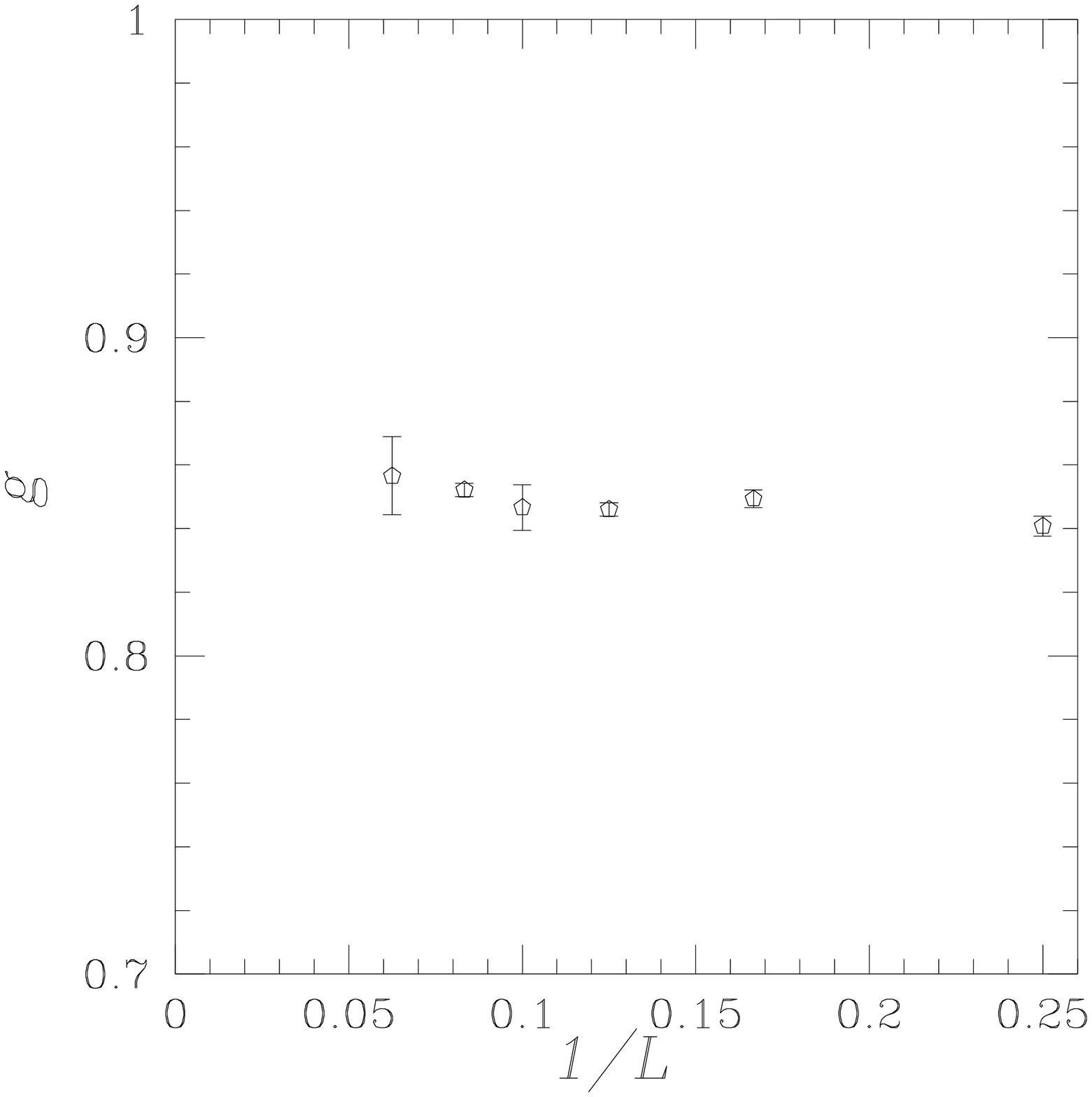}
\end{center}
\caption[0]{\protect\label{fig:binder07}
The overlap Binder cumulant at fixed $T=0.7$ versus $1/L$.}
\end{figure}

These results support and improve the results of \cite{KAWYOU}, where
the crossing was first observed.  Here we offer evidence for a split
of the curves in the low $T$-region down to $T\simeq 0.7T_{c}$
(improving over the single point at $T\simeq 0.9T_{c}$ of
\cite{KAWYOU}).  We notice that the splitting is much smaller in the
spin glasses than in usual ferromagnets because also in the infinite
volume limit, if replica symmetry is broken, the Binder cumulant is
not identically equal to one in the low temperature phase, but it is
given by
\be
g=\frac32- \frac12{\int \d q~P(q)q^{4} \over (\int \d
q~P(q)q^{2})^{2}}\ .
\ee
The small dependence of $g$ on $L$ reflects the mild dependence of
$P(q)$ on the size.  It is evident that $g$ does not go to 1 at
$T<T_{c}$ (see fig. (\ref{fig:binder07})) implying that the function
$P(q)$ does not tend to a delta function in the infinite volume limit
in contradiction with the droplet model predictions \cite{DROPLET}.

 The second information that we can obtain from the analysis of the
Binder cumulant is the value of the $\nu$ critical exponent.  We have
estimated the value of the exponent $\nu$ by fitting the derivative of
$g$ in the high temperature region with the technique introduced in
\cite{INPARU}.  We introduce the function $f$ as the derivative of the
Binder cumulant $g$ with respect to $T$ at fixed value of $g=g_{0}$:

\begin{figure}[htbp]
\begin{center}
\leavevmode
\epsfysize=250pt
\epsffile{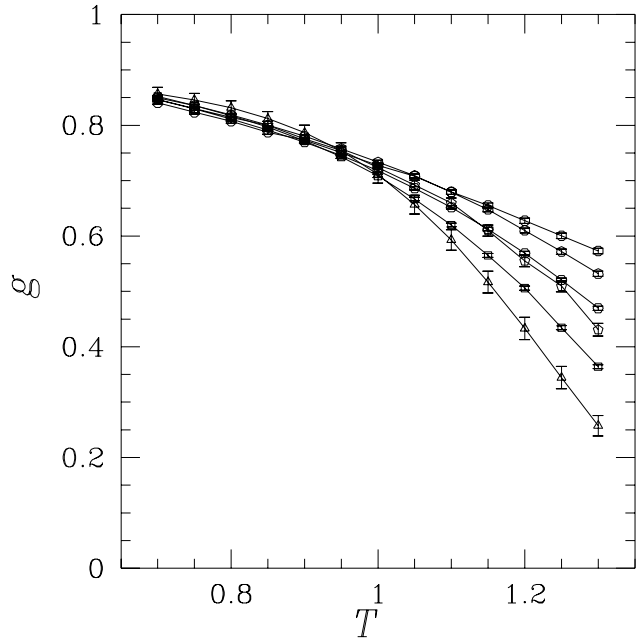} 
\end{center}
\caption[0]{\protect\label{fig:binder1}
The overlap Binder cumulant versus $T$ for $L=4,6,8,10,12$ and $16$ 
(curves from top to bottom on the right part of the figure).}
\end{figure}

\begin{figure}[htbp]
\begin{center}
\leavevmode
\epsfysize=250pt
\epsffile{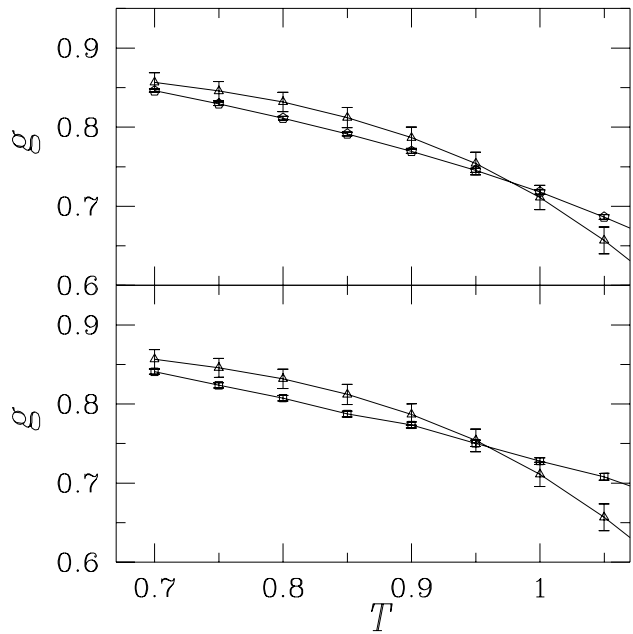}
\end{center}
\caption[0]{\protect\label{fig:binder2}
A magnified view of figure (\ref{fig:binder1}). 
In the upper window crossing between the $L=8$ and $L=16$
lattices. In the lower window crossing between 
$L=4$ and the $L=16$ lattices.}
\end{figure}

\be
f(g_0,L)\equiv\left. \frac{\d g}{\d T}\right|_{T_0:\; g(T_0)=g_0}
\simeq \alpha L^{\frac{1}{\nu}} \ .
\protect\label{nu_alpha}
\ee
We have computed the derivatives of the overlap Binder cumulant in the range 
from $g=0.6$ to $g=0.8$ (the crossing point is close to the value $0.75$).  
In this interval we have fitted the Binder cumulant curves (at fixed 
$L$) using fourth and fifth order polynomials (in order to control the 
systematic effects due to the fitting procedure).  To cross check 
consistency we have also used a derivative both with respect to the 
temperature and with respect to
$\beta$: the two procedures are equivalent and they should produce the 
same result.

\begin{figure}[htbp]
\begin{center}
\leavevmode
\epsfysize=250pt
\epsffile{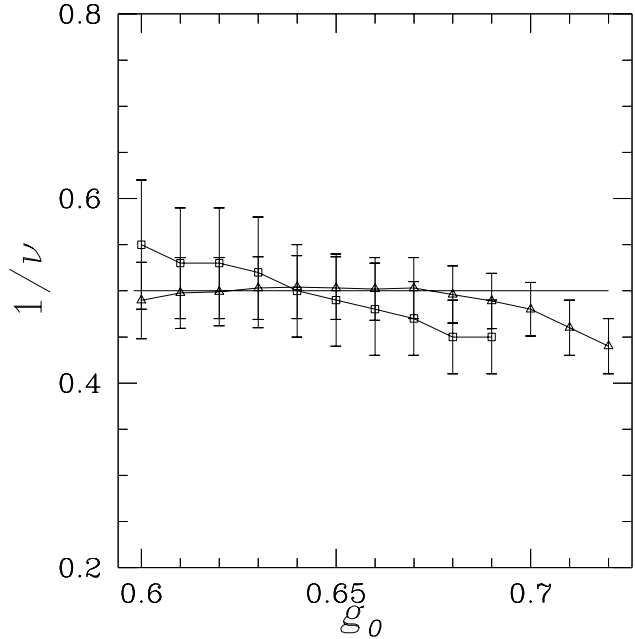}
\end{center}
\caption[0]{\protect\label{fig:binder_nu}
$\frac{1}{\nu}(g_{0})$ versus $g_{0}$. 
Triangles for fits over $T$, squares for fits over $\beta$.}
\end{figure}

In a large region with $T>T_{c}$ close to the critical point such an 
estimate turns out to be, as hoped and expected, independent of the 
$g_{0}$ value that has been used.  We plot in figure 
(\ref{fig:binder_nu}) the values of $\frac{1}{\nu}(g_{0})$ as a 
function of $g_0$ (obtained fitting the data with a fourth degree 
polynomial both in $T$ and in $\beta$). We expect the fit to break down 
when going too close to $g_{c}$ because of the large error involved.

The fit to the power behavior of eq.  (\ref{nu_alpha}) works well in 
the range of $g_{0}$ going from $g_0=0.6$ to $g_0=0.68$ (i.e the 
$\chi^2/{\rm DF} \simeq 1$, DF stands for degrees of freedom): 
for $g_{0}\ge 0.69$ we get to close to 
the crossing and the fit becomes unfaithful. This happens 
both for the $\beta$ and the $T$ fit. The error looks bigger in the 
$\beta$ interpolation.  The final value is

\be
  \nu=2.00 \pm 0.15 \ .
  \protect\label{E-NUA}
\ee
Let us stress the ``locality'' of this method. We perform the
derivative closer and closer the critical point. When we determine a
plateau we know we are entering the critical region where scaling
violations are small. 
In reference \cite{KAWYOU}, Kawashima and Young simulated the $\pm 1$
version of the model using an annealing procedure in order to
thermalize the system. The larger lattice that they simulated was
$L=16$.  They found $\nu=2.0$ analyzing the Binder cumulants curves.
Another value of $\nu$ was obtained analyzing the scaling of $P(q)$
near the critical temperature, $\nu=1.6$. (obtained also $\beta/\nu$).
Finally they reported as final value $\nu=1.7(3)$.

Instead of study the scaling properties of the overlap probability
distribution we will follow another way. We will compute, using the
Binder cumulant data, the infinite volume critical temperature, and
using the power law scaling of the susceptibility at the critical
point we will get the critical exponent $\gamma/\nu$.

The Binder cumulant also gives us quantitative information about the 
value of the critical temperature.  One can see \cite{PARU} that the 
temperature where the Binder cumulant takes a given value, that we 
call $g_{1}\equiv g(L,T_c(L,g_1))$, close enough to $g_{c}$, scales as
\be
T_c(L,g_1)=T_c(\infty) + a L^{-1/\nu} \ .
\protect\label{binder_fit}
\ee
Such a value of the infinite volume critical temperature and of the 
critical exponent will depend on $g_1$ if it is not close enough to 
$g_{1}$: when taking the infinite volume limit we have to keep $g_{1}$ 
close enough to $g_{c}$.  Inside the critical region we will find that 
asymptotically the critical temperature and the critical exponent 
become independent of $g_1$. For instance the three parameter fit 
using $g_{1}=0.68$ give us

\be 
  T_c=0.99 \pm 0.07\  \ .  
  \protect\label{E-NUB}
\ee 
Nevertheless the determination of $\nu$ has a very large error.  
In order to reduce the eror bars in $T_c$ we fix $\nu$ to the value 
given in Eq. (\ref{E-NUA}): $\nu=2.00(15)$
Fixing  $\nu=2$ in 
(\ref{binder_fit}) we find 
$T_c=0.92(3)$ for $g_1=0.65$; 
$T_c=0.95(4)$ for $g_1=0.68$ and $T_c=0.92(3)$ for 
$g_1=0.70$. For $g_{1}>0.70$  the quality of the fit is poor.  
Taking into account all these numerical results and the $\nu$'s error 
bars~\cite{F1} we finally 
quote the result

\be
T_c=0.95\pm 0.04\ , \, \, \nu=2.00\pm 0.15 \ .
\ee

\begin{figure}[htbp]
\begin{center}
\leavevmode
\epsfysize=250pt
\epsffile{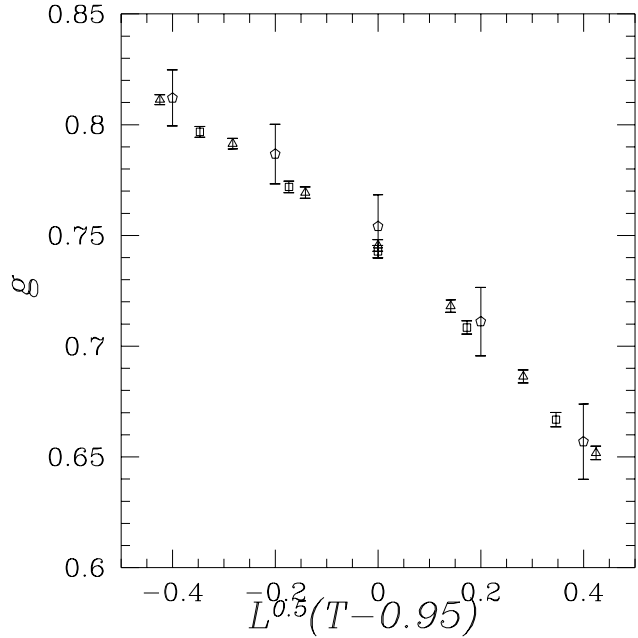}
\end{center}
\caption[0]{\protect\label{fig:scaling_binder}
$g$  versus 
$L^{\frac{1}{\nu}}(T-T_c)$, with
$\nu=2.0$ and $T_c=0.95$. In the precision given by the statistical
error all the points (from different
lattice sizes) collapse on the same curve. Triangles for $L=8$, 
squares for $L=12$ and pentagons for $L=16$.}
\end{figure}

We end this quantitative study with the standard plot of the overlap
Binder cumulant, in figure (\ref{fig:scaling_binder}): we plot $g$
versus the scaling variable $L^{1/\nu} (T-T_c)$, using $\nu=2.0$ and
$T_c=0.95$. All the points (from different lattices and temperatures)
go to  the same curve giving us a very good scaling plot.

After locating with  precision 
the location of the critical point we can study the divergence
of the spin glass susceptibility. 
The susceptibility diverges with a 
power law with an exponent given by $\frac{\gamma}{\nu}$.  Fitting our 
data at $T_c=0.95$ we obtain $\frac{\gamma}{\nu}=2.36$.  We must take 
into account the error bars on the critical 
temperature. At $T=0.95+0.04$ we find 
$\frac{\gamma}{\nu}=2.30$ while at $T=0.95-0.04$ we have 
$\frac{\gamma}{\nu}=2.42$, so we quote

\be 
  \frac{\gamma}{\nu}=2.36\pm 0.06 \ .
\ee
These exponents  are in good agreement with those found by
Kawashima and Young \cite{KAWYOU} for the $J =\pm 1$ Ising spin glass,
$\nu=1.7 \pm 0.3$ and $\frac{\gamma}{\nu}=2.35\pm 0.05$, and with the 
$\frac{\gamma}{\nu}=2.37\pm 0.04$ found by Berg and Janke in 
\cite{BERJAN}.
\section{\protect\label{S_PQ}Analyzing $P(q)$}

The issue of the properties of the probability distribution of the
overlap is controversial. The $P(q)$ is a crucial ingredient of the
RSB picture: its equilibrium non trivial shape implies a whole series
of unusual patterns of behaviors.

In the droplet model \cite{DROPLET} $P(q)$ has very
simple properties: it is composed by two Dirac deltas at $\pm q_{\rm
EA}$, where $q_{\rm EA}$ is the Edward-Anderson order parameter.  

Moreover we can cite the approach of reference \cite{NS} about
another definition of the overlap (the so-called window overlap)
which must have a droplet model like  probability distribution. 

However in reference \cite{WINDOW} has been shown that both overlaps,
the first one computed in all the lattice and the second one, the
window overlap, computed in a small region around the origin
(following the prescriptions of reference \cite{NS}), provide the same
picture of the probability distribution $P(q)$, supporting both
definitions of the overlap the RSB picture.

\begin{figure}[htbp]
\leavevmode
\epsfysize=250pt
\centerline{\epsffile{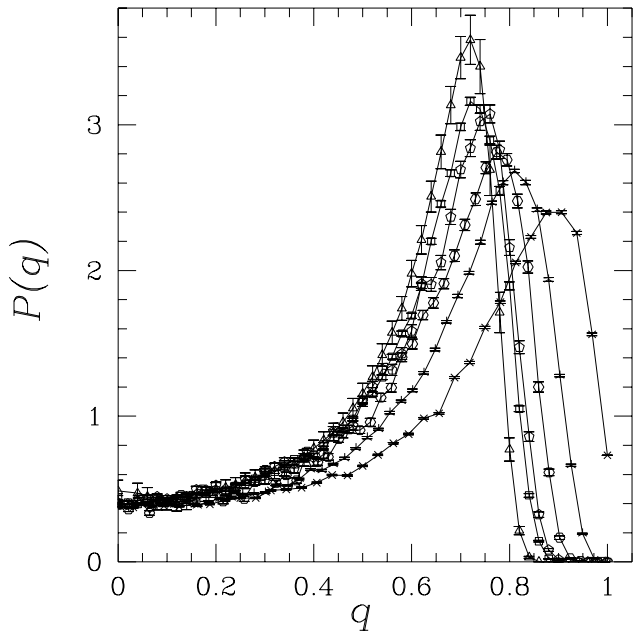}}
\caption[0]{$P(q)$ probability distributions at $T=0.7$ for all the 
simulated lattice sizes (right to left):4, 6, 8, 10, 12 and 16.}
\protect\label{fig:figura2}
\end{figure}

We will show below, that $P(q)$ does not follow the shape predicted by
the droplet model \cite{DROPLET}, following qualitatively the
properties predicted by Mean Field.
The average properties of the function $P(q)$ have been discussed in
reference \cite{INMAPARU}.  The results are shown at the lowest
temperature in figure (\ref{fig:figura2}).  The position of the peak
($q_M$) was fitted as (at the lowest temperature we considered,
$T=0.7$)

\be
q_M = (0.70 \pm 0.02) + (1.6 \pm 0.7) L^{-(1.5 \pm 0.4)}\ .
\ee
Let us discuss 
how the system develops with increasing $L$ a Dirac delta function at
$q=q_{EA} \simeq 0.7$. We define

\begin{equation}
  A(L) \equiv \int_{q_{\rm max}} ^1 \ d|q| P_L(|q|) \ ,
\end{equation}
where $q_{\rm max}(L)$ is the overlap value where $P_L(|q|)$ is
maximum.  If in the infinite volume limit the system develops a Dirac
delta function at $q_{EA}=0.7$ $A(L)$ should not depend on $L$ (at
least for $L$ large enough).  The weight of the Dirac delta will be $2
A$ (if the delta is
approximated by a symmetric function).

For instance at $T=0.7$ we find 
$A(4) =0.25\pm 0.03$, 
$A(6) =0.27\pm 0.01$,
$A(8) =0.26\pm 0.01$, 
$A(10)=0.22\pm 0.05$, 
$A(12)=0.27\pm 0.03$ and finally
$A(16)=0.25\pm 0.04$. 
We can assume that the previous values are roughly
constant ($A \simeq 0.26\pm 0.02$), and we expect that the weight
of the Dirac delta at $q_{EA}=0.7$ will be, if the delta is
approximated by a symmetric function,  $0.52\pm 0.04$.
This implies that 
it is consistent to assume that the peak will become a delta
function in the infinite volume limit.  
In any case the weight must be substantially larger than $0.26$.

Moreover this result supports the existence of a continuous part of
$P(q)$ between $0$ and $q_{EA}=0.7$, that carries the missing probability
$0.48$ (if the delta is
approximated by a (almost) symmetric function), 
in order to make $\int_0^1 \ d |q| P(|q|)= 1$.  This analysis
gives further support to the existence of a replica symmetry breaking
phase transition in the $3D$ EA spin glass.

\begin{figure}[htbp]
\begin{center}
\leavevmode
\epsfysize=250pt
\centerline{\epsffile{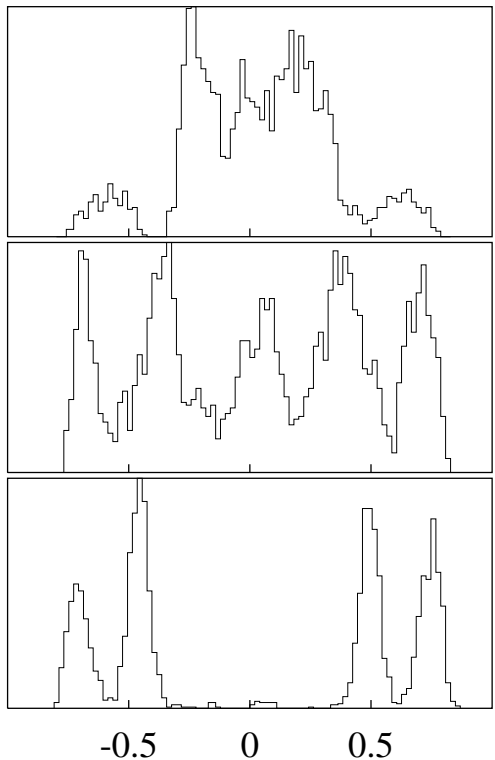}}
\end{center}
\caption[0]{\protect\label{fig:trepq}
Non normalized $P_{J}(q)$ (histogram) 
for three different realizations of the quenched disorder,
$L=16$ and $T=0.7$.}
\end{figure}

According to the replica predictions the function $P_{J}(q)$ changes
dramatically from sample to sample.  A smooth dependence on $q$ is
obtained only after average.  We show in figure (\ref{fig:trepq}) three
different functions $P_{J}(q)$ for three different samples.
One can see that different samples have a very different behavior. In 
the first distribution on the top we have a sizable contribution 
for $q\simeq 0$: these kind of samples contribute to the low $q$ part 
of $P(q)$ (exactly like it happens in mean field: only some samples 
carry weight at $q\simeq 0$, while some samples have a zero 
probability for $q\simeq 0$ overlaps). The sample in the middle has 
five clear maxima, one in $q=0$ and four at high $q$ values. The 
lower sample has $4$ very sharp maxima (here the similarity with a 
$\delta$ function starts to be clear), and close to no weight at 
$q\simeq 0$. Notice that the level of symmetry of these plots for 
$q  \leftrightarrow -q$ is a good check of how good the thermalization 
has been.

Here we will compare in quantitative way the detailed predictions of
the sample to sample fluctuations of the function $P_{J}(q)$ with the
very detailed predictions of replica theory \cite{MPSTV,MPVFRE}, and
we will find a remarkable agreement.

From the $P_{J}(q)$ of (\ref{E-PQJ}) we define the integrated
probability distribution

\be
  x_J(q_s)
  \equiv \int_0^{q_s} \d q\ P_J(q)
  \equiv 1-y_J(q_s) \ ,
\ee
i.e.  the total probability that the overlap in a given sample is 
smaller than $q_s$.  In the following $q_s$ will be kept fixed.  The 
probability distribution of $x_J(q_s)$, that following \cite{MPVFRE} 
we denote by $\hat{\Pi}_{q_s}(x_J)$, is also a random variable, and it 
depends on the disorder realization $J$.  In order to get a better 
statistical signal (for reasons that will become clearer in the 
following) it is useful to define the integrated probability of 
$\hat{\Pi}_{q_s}(x_J)$

\be
  \hat{\Pi}_{q_s}^<(r)=\int^r_0 \d z ~ \hat{\Pi}_{q_s}(z) \ .
\ee
We define also

\be
x(q_s)\equiv \overl{x_J(q_s)}=\int_0^{q_s} \d q ~ P(q)\equiv 1-y(q_s)  \ . 
\protect\label{x_def}
\ee
With our numerical simulations we compute $P_J(q)$ for every 
realization of the quenched disorder $J$.  We select a fixed value of 
$q_s$, and for each realization $J$ we compute the probability to find 
at equilibrium an overlap smaller than $q_s$.  The answer to the 
question {\em ``for which percentage of disorder realizations such a 
probability is smaller than $r$?''} is given by 
$\hat{\Pi}_{q_s}^<(r)$.  This procedure is repeated for a set of 
$q_{s}$ values (in the numerical work we have used $q_s$ $=$ $0.1$, 
$0.2$, $\ldots$, $0.9$, see below).  

The properties of this distribution have been carefully studied in the
past \cite{MPSTV,MPVFRE} in the framework of the Mean Field RSB
theory.  An analytic expression can be obtained for it; indeed one can
prove that \cite{MPVFRE}

\be
  \int_0^\infty \d v\ e^{-z v} g_{q_s}(v)= 
  \frac{1-x(q_s)}{z} 
  \frac{
    D_{x(q_s)-2}
    \left(\frac{-1}{\sqrt{2 z}}\right)
  }{
    D_{x(q_s)}
    \left(\frac{-1}{\sqrt{2 z}}\right)
  } \ ,
  \protect\label{E-SOLUTI}
\ee
where $\Pi_{q_s}(y_J) \equiv \hat{\Pi}_{q_s}(1-x_J)$ (i.e.  it is the 
probability distribution of the random variable $y_J(q_s)$), and

\be
  g_{q_s}(u) \equiv \int_0^1 \d z~ \Pi_{q_s}(z) e^{\sqrt{u/z}} \ .
\ee
The $D_\alpha$ are the parabolic cylindric functions,
and  $x(q_s)$ has been defined in (\ref{x_def}). 

The function $\Pi_{q_s}(z)$ has many interesting properties.  We will
concentrate on some of the asymptotic properties of $\Pi_{q_s}(z)$
for $z$ close to 1.  It has been proven that the previous formulae imply
that:

\be
  \Pi_{q_s}(z) \simeq (1-z)^{x(q_s)-1}\, \, \, \,\,\, 
  {\rm as}\, \, z  \to 1
  \ ,
\ee
or equivalently

\be
\hat{\Pi}_{q_s}(z) \simeq z^{x(q_s)-1}\, \, \, 
\,\,\, {\rm as} \, \, z \to 0 \ .
\ee
Finally, in terms of the integrated probability distribution

\begin{equation}
  \hat{\Pi}_{q_s}^<(z) \simeq z^{x(q_s)} \, \, \, \,\,\, 
  {\rm as} \, \, z \to 0 \ .
\end{equation}
The integrated probability distribution from 0 to $z$ (for a fixed 
reference value $q_s$) goes to zero (when $z \to 0$) as a power law of 
$z$ with exponent $x(q_s)$.  What we are looking for is the 
probability of finding a function $P_{J}(q)$ which is very small in 
the region $0<q<q_{s}$ (i.e. $\int_{0}^{q_{s}} dq P_{J}(q)<z$), and 
this probability goes to zero as a power of $z$.  This behavior is 
quite peculiar, especially in the region where $x(q_s)$ is small.  
Indeed we find that $\hat{\Pi}_{q_s}^<(z)$ remains a quantity of order 
$1$ as far as we stay in the region

\be
z >  e^{-\frac{1}{x(q_s)}}\ .
\ee
When $x(q_s)$ is small the probability distribution of $z$ becomes very {\sl
intermittent}

\bea
\overline{z}&=& x(q_s)\\
e^{\overline {\log(z)})} &=& e^{-\frac{1}{x(q_s)}}\ .
\eea
Where $\log x$ means, in the whole paper,  the natural  logarithm of $x$.
Before using our numerical simulations of the $3D$ model to check if
we find a similar power behavior, it is convenient to look more
closely to the analytic predictions.  In the RSB solution of the mean
field theory one can directly inspect the solution (\ref{E-SOLUTI}) to
exhibit the power law behavior.  However the computation of the
function itself is rather lengthy because it involve the evaluations
of inverse Laplace transforms.  Let us illustrate here a different way
to obtain the correct answer, which is based on probabilistic
techniques and has also the advantage of giving a further insight in
the features of the RSB solution.

\begin{figure}[htbp]
\begin{center}
\leavevmode
\epsfysize=250pt
\epsffile{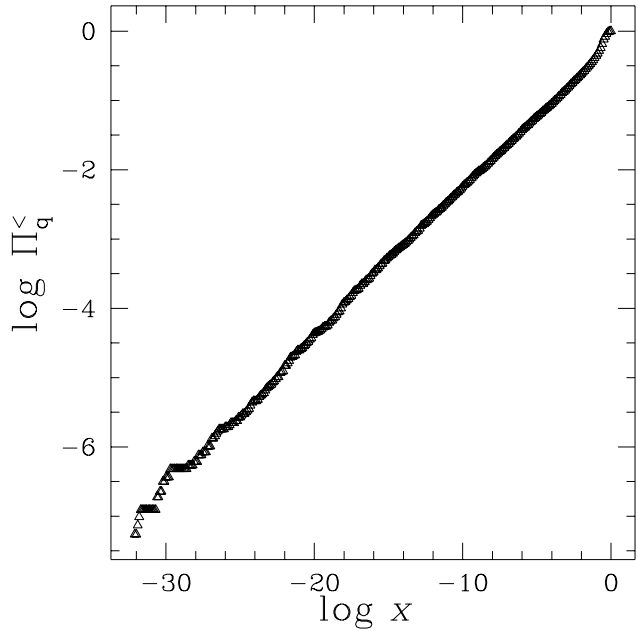}
\end{center}
\caption[0]{
  \protect\label{fig:ris}
$\log(\Pi_q^<(x))$ versus $\log(x)$ in the mean field approximation 
(log means the natural logaritm). Here $q$ has been fixed to a value 
such that $x_{c}=0.2$ (see text).  
The clear power law behavior is well fitted by an exponent equal to 
$x_{c}$.}
\end{figure}

We recall that in the RSB theory the ultrametric structure of the
states implies that it is possible, by fixing a given value of the
overlap $q_c$, to break the set of all the pure states in clusters
\cite{MPSTV}.  Two pure states belong to the same cluster if their
mutual overlap is bigger than $q_c$.  This is equivalent to fix a
value of $x_c\equiv x(q_c)$, since the relation between $q$ and $x$ is
monotonic (see equation (\ref{x_def})).  Each cluster is characterized
by a weight $w_I^J$, where $I$ is the index of the cluster and $J$
means that we are working in a given disorder realization, that is
nothing but the sum of the weights of all the pure states belonging to
the cluster.  The probability of finding an overlap greater than $q_c$
(the reference value in the construction of the clusters) turns out to
be
\be
  y_J(q_c)\equiv 1-x_J(q_c)= \sum_I (w_I^J)^2 \ ,
  \protect\label{weight_y}
\ee
where the index $I$ is running over all the 
  clusters~\cite{F2}.

A theoretical analysis tells us that the probability distribution of
the weights of the clusters can be generated by the following
algorithm.  We consider a system where $N$ clusters are allowed
(eventually $N$ will go to infinity).  We assign the weights $w_I$ to
the $N$ clusters as follows
 
\be
  w_I = A e^{-\beta F_I}\ ,
\ee
where $F_I$ is the extensive free energy of cluster $I$, and $A$ is a
($F$-dependent) normalization constant such that
$\sum_{I=1,N}w_{I}=1$.  The extensive free energies are mutually 
independent and they are distributed according to

\begin{equation}
  P(F_I)=\left\{
\begin{tabular}{c c c}
$ B e^{\beta x(q_c) F_I}$& if &  $F<F_{M}$\ ,\\
$0$ &  & otherwise \ .\\
\end{tabular}
\right.
\end{equation}
where $B$ is a normalization factor.  With these weights we can
construct the quantity $1-\sum_I w_I^2$ and compute the probability
distribution of these sums (the $\hat{\Pi}_{q_c}$) or the integrated
probability distribution $\hat{\Pi}^<_{q_c}(r)$.  The result for the
weights does not depend on $F_{M}$, which we can take equal to 0.  If
we use this algorithm in the limit where $N \to \infty$ we obtain the
correct probability distribution.

This algorithm can be easily implemented numerically.  The probability
distribution of of the quantity
$\omega_{I}\equiv \exp({-\beta F_I})$ is given by
\be
  P_{q_c}(\omega) \propto \frac{1}{\omega^{x(q_c)+1}} 
  \protect\label{weight}
\ee
can be now generated by extracting $N$ weights $w$ using the random
numbers $p$, uniformly distributed in $(0,1)$:
$w=A\frac{1}{p^{1/x_c}}$.  With these weights we can construct the
quantity $1-\sum_I w_I^2$ and compute the probability distribution of
these sums (the $\hat{\Pi}_{q_c}$) or the integrated probability
distribution $\hat{\Pi}^<_{q_c}(r)$.

Very good results can be obtained also for moderate values of $N$
(i.e.  $N>10$ if we do not consider a value of $x$ close to $1$.  We
show in in figure (\ref{fig:ris}) the result of this algorithm for a
choice of $q_{c}$ such that $x_c=0.20$.  As it should be the data
follow a very accurate power law in the range of $\log(x)$ (natural log) 
going to
$-25$ to $-1$, with an exponent of $0.2$.

\begin{figure}[htbp]
\begin{center}
\leavevmode
\epsfysize=250pt
\epsffile{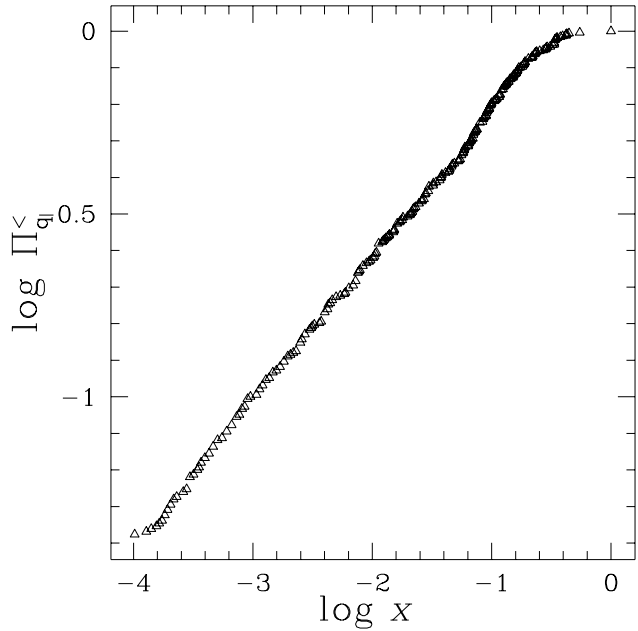}
\end{center}
\caption[0]{
  \protect\label{fig:ris_3d}
  $\log \Pi_q^<(x)$ versus $\log x$ (log means natural logarithm)
  from the results of numerical simulations of the $3D$ Edwards Anderson 
  spin glass. Here $q=0.3$ (which
  corresponds to $x_c=x(0.3)\simeq 0.15$),  $L=16$ and $T=0.8$.
}
\end{figure}

\begin{figure}[htbp]
\begin{center}
\leavevmode
\epsfysize=250pt
\epsffile{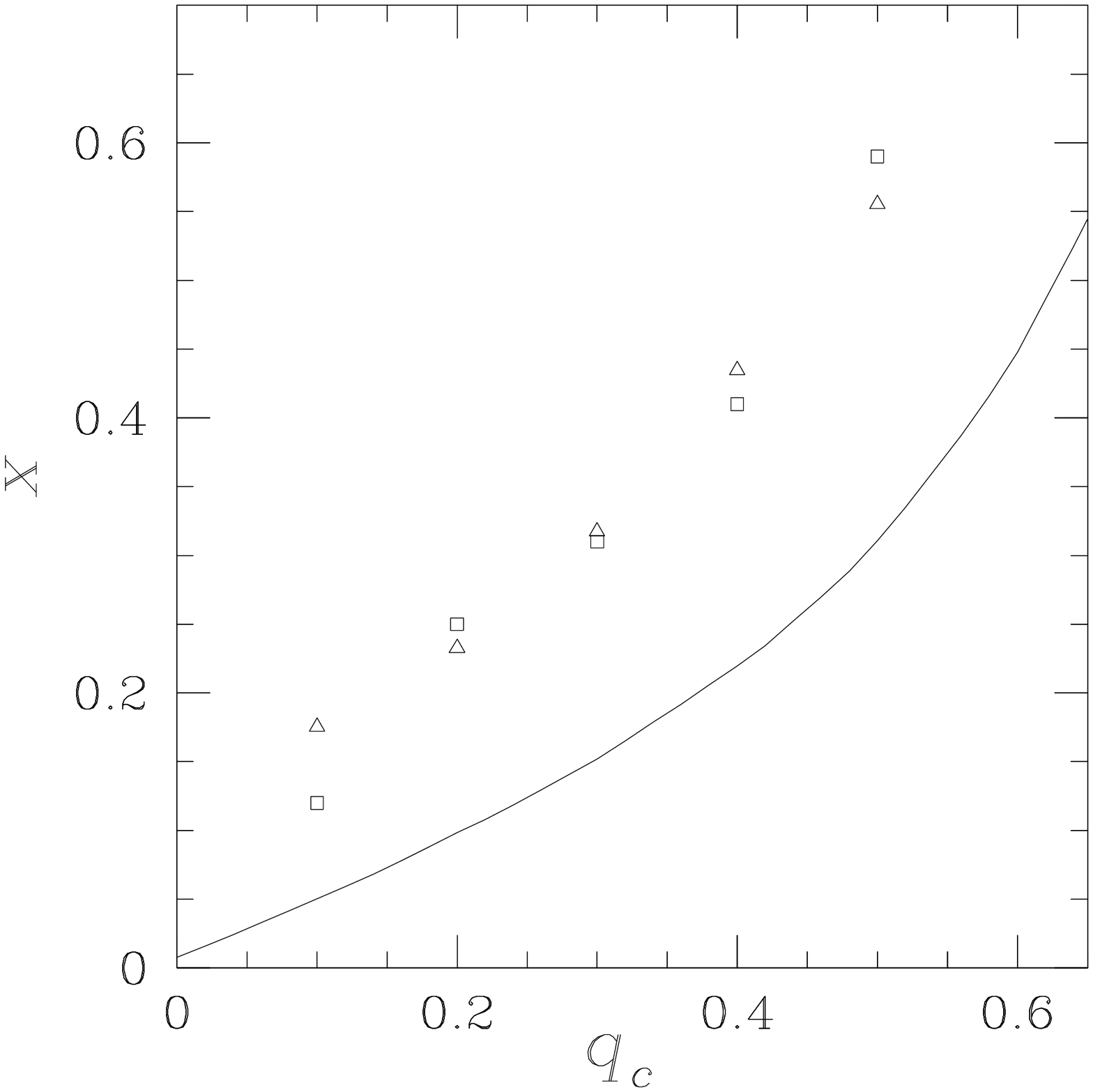}
\end{center}
\caption[0]{
  \protect\label{fig:x}
  The exponent $x_{3D}(q_{c})$ obtained on a $L=10$ lattice (squares) 
  and on a $L=16$ lattice (triangles) at  $T=0.8$.  The continuous line 
  is the equilibrium function $x(q)$, obtained by integrating the 
  equilibrium probability function, $P(q)$, using equilibrium data on 
  a $L=12$ lattice.
}
\end{figure}

We plot the results for $\hat{\Pi}_{q_c}^<(r)$ from the numerical 
simulations of the $3D$ model in figure (\ref{fig:ris_3d}).  The power 
law behavior is clear, and consistent with the results of mean field 
theory.  The exponent $x_{3D}(q_{c})$ does not coincide exactly, as a 
function of $q_{c}$, with the function $x(q_c)$ (calculated by 
integrating directly the equilibrium $P(q)$): we plot
the two functions in figure (\ref{fig:x}).

\begin{figure}[htbp]
\begin{center}
\leavevmode
\epsfysize=250pt
\epsffile{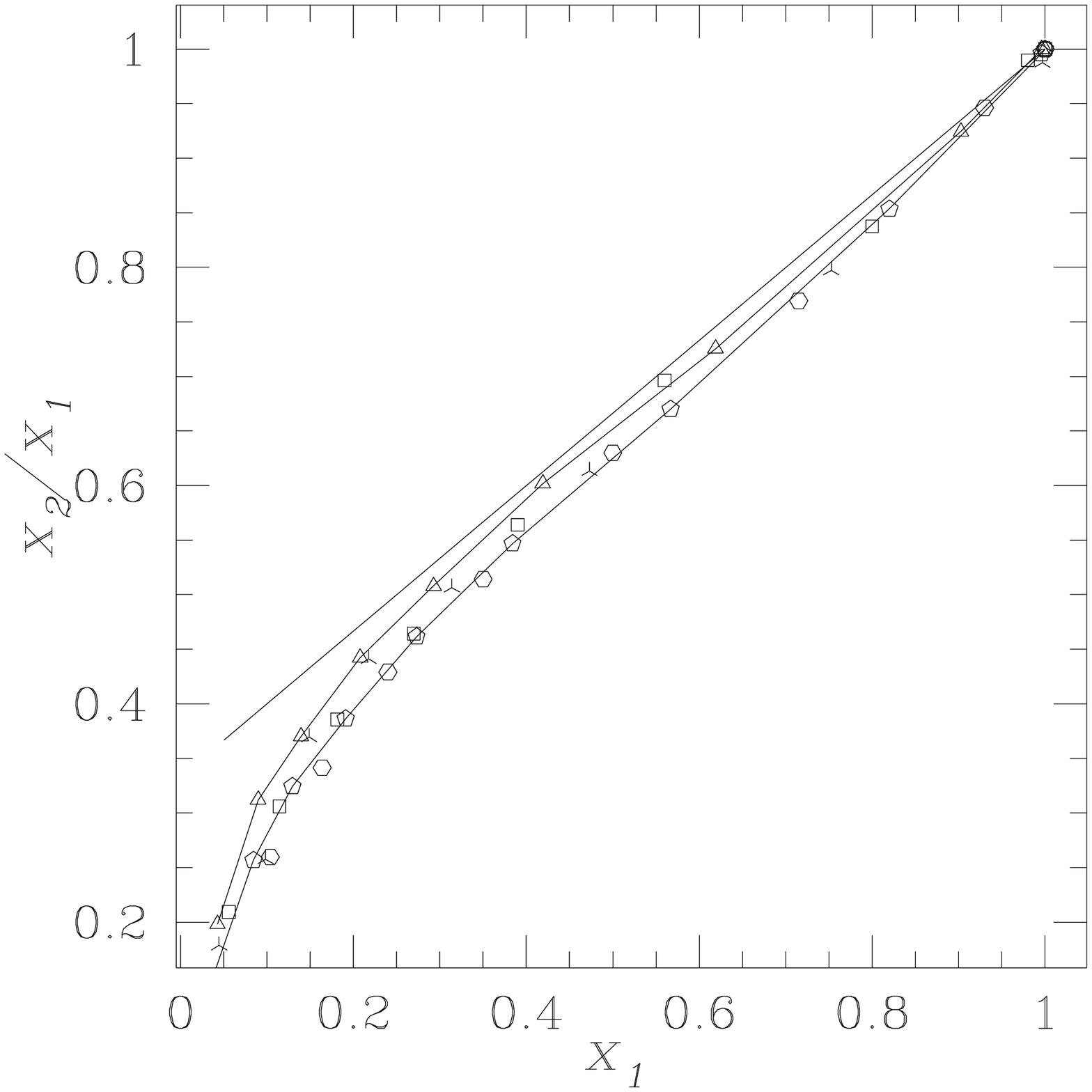}
\end{center}
\caption[0]{
  \protect\label{fig:x2_x1}
  $\frac{x_2}{x_1}$ versus $x_1$.  For $L=10$ $T=0.7$ (triangles), and 
  $T=0.8$ (squares);  for $L=8$ $T=0.7$ (pentagons), and $T=0.8$ 
  (hexagons); for $L=16$ at $T=0.7$ (heptagons)  
   Triangles and pentagons are connected with continuous 
  lines.  The straight line is the mean field prediction.
}
\end{figure}

We can obtain more insight about $P(q)$ by analyzing the moments of the 
$\hat{\Pi}_{q_c}(r)$ probability distribution. We consider the 
relations \cite{MPSTV}, valid in the RSB solution of the mean field 
model,

\begin{eqnarray}
  x_2(q_c)&=&\frac{1}{3} x_1(q_c) + \frac{2}{3} x_1^2(q_c)\ ,
  \protect\label{E-X123}\\
  x_3(q_c)&=&\frac{x_1(q_c)}{15} \left(3 +7 x_1(q_c) +5
  x_1(q_c)^2\right) \ ,
\protect\label{E-XX}
\end{eqnarray}
where 

\be
  x_n(q_c)\equiv \int \d r\  r^n \hat{\Pi}_{q_c}(r) \ .
\ee
We plot in figure (\ref{fig:x2_x1}) $\frac{x_2}{x_1}$ versus $x_1$ for
the $L=8, 10$ and the $L=16$ lattices at two different temperatures:
$T=0.8$ and $T=0.7$.  We also show the straight line prediction of
mean field theory (\ref{E-X123}). We note the numerical data do not 
depend on the temperature.
The figure also shows that for $x>0.5$ there is a 
good agreement among the mean field RSB solution and the numerical 
data for the $3D$ model.  For $x<0.5$ data on larger lattices are 
becoming closer to the mean field result.  Recent analytic and 
rigorous work \cite{GUERRA,AIZCON,PARSIX} is focusing on relations of 
the type (\ref{E-X123}), and on the relations among expectation values 
with different number of replicas.

\begin{figure}[htbp]
\begin{center}
\leavevmode
\epsfysize=250pt
\epsffile{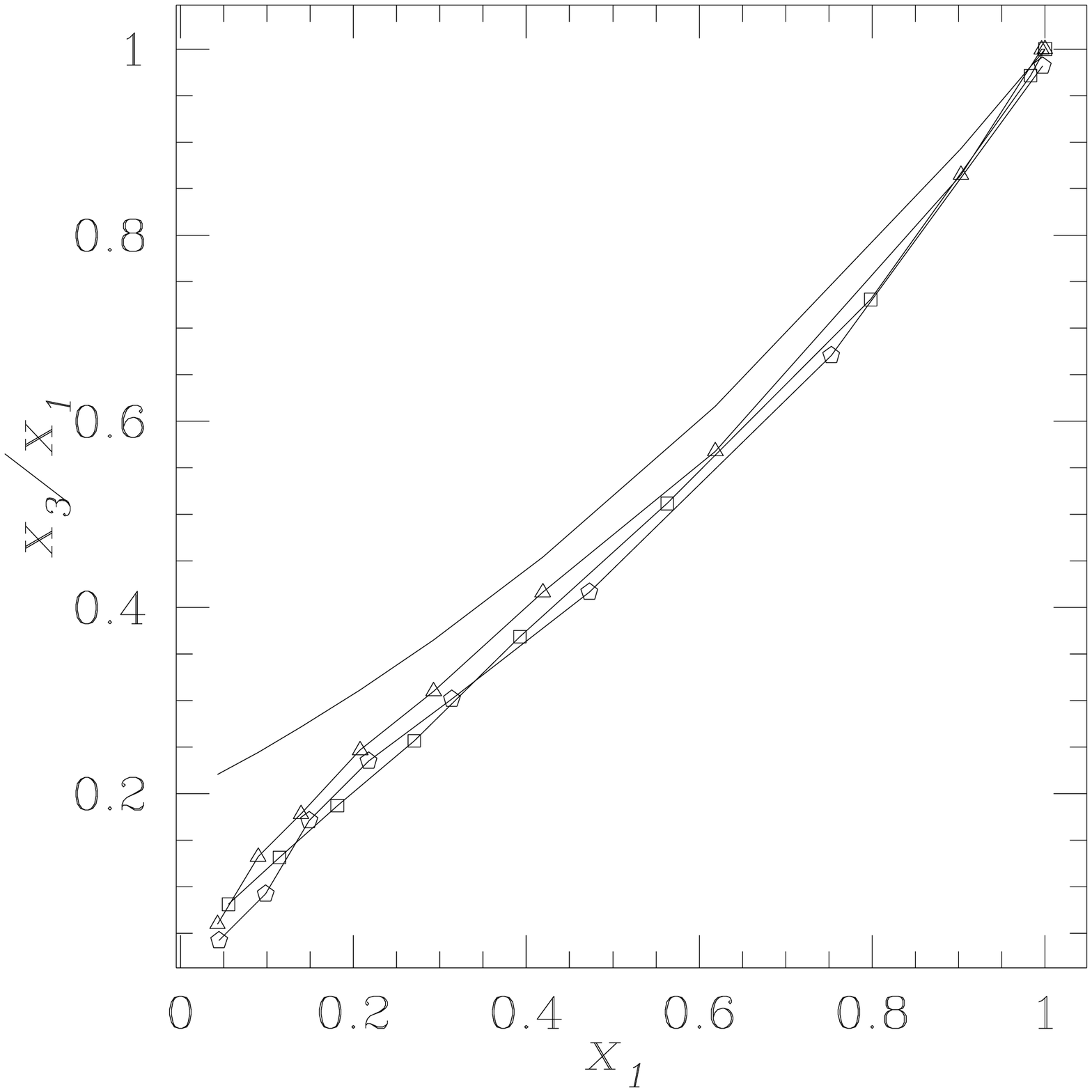}
\end{center}
\caption[0]{
  \protect\label{fig:x3_x1}
  $\frac{x_3}{x_1}$ versus $x_1$.  For $L=10$ $T=0.7$ (triangles), and 
  $T=0.8$ (squares).  For $L=10$ $T=0.7$ (pentagons).    The upper
  continuous  line is the mean field prediction.
}
\end{figure}

We have obtained similar results analyzing $x_3/x_1$ against $x_1$. Our
numerical data are plotted in figure (\ref{fig:x3_x1}) 
together with
the Mean Field prediction of equation (\ref{E-XX})). 

We have seen that the replica predictions are in good agreement 
with the numerical data. There are 
however two problems:

\begin{itemize}
\item
the relations (\ref{E-X123}) and (\ref{E-XX}) seems not 
to work at small $x_1$;

\item
the value of the dynamical exponent $x_{3D}(q_{c})$ is higher that
what it should be.
\end{itemize}
It is not clear from our data if these discrepancies are a finite
volume effects or if they disappear when the volume goes to infinity.
We want to point out that these discrepancies are just what we would
expect from finite size effects.  Indeed in a finite volume we could
expect that

\be
P(q)=\sum_{\alpha,\beta}w_{\alpha}w_{\beta} 
f(q-q_{\alpha,\beta},\Delta_{\alpha,\beta})\ ,
\ee
where the function $f(z,\Delta)$ is essentially different from zero
for $|z|< \Delta$. In the limit $\Delta \to 0$ we should have
$F(z,\Delta) \to \delta(z)$.  It is obvious that delta function
only appears in the infinite volume limit and the quantity $\Delta$ is likely
to go to zero in the average as a power of $L$.  In this situation the
quantity $x_{J}$ is always of order $q$ when $q$ goes to zero (the
function $P_{J}(q)$ being bounded) and therefore the expectation of
$x_{J}^{2}$ is of order $q^{2}$.  Consequently in the region where
$q<\Delta$ we must have that the ratio $\frac{x_2}{x_1}$ must go to
zero with $x$.  It is remarkable that this behavior sets in
only at small values of $x_{1}$.

The same argument can be applied to the tails of 
the probability distribution.
For sake of simplicity we consider the case where

\be
f(z,\Delta)={\theta(\Delta^{2}-z^{2})\over 2 \Delta} .
\ee
It is evident that the the value of $P(q)$ will be affected by the
presence of states having an average overlap different at most
$\Delta$ from $q$.  A simple computation show that in the
approximation where $\Delta$ does not depend on the states $\alpha$
and $\beta$, we have that

\be
x_{3D}(q_{c})=x(q-\Delta),
\ee
which is roughly what we observe, with $\Delta \approx 0.15$. This
value is of the same order of magnitude of the width of the function
$P(q)$ at the peak and also is of the same order of magnitude of the
region, in $x_1$, in which the relations (\ref{E-X123}) and
(\ref{E-XX}) do not work. A much more accurate analysis is needed to
decide if these small discrepancies arise from the mechanism outlined
here and consequently disappear at large volumes.

\section{\protect\label{S_SUM}A Simple Function}

Let us consider the function

\begin{equation}
\beta (1-\overline{\lan |q| \ran})\ ,
\end{equation}
that we will call $F$. In the RSB solution of the mean field model one
can compute its behavior.  Usual arguments imply that $F=\chi$, $\chi$
being the magnetic susceptibility.  In the high temperature phase
$P(q)$ is a delta function, and $\overline{\lan |q| \ran}=0$:
everybody agrees that this result also holds in finite dimensional,
realistic spin glasses.

In the SK model one finds that
\be
  \overline{\lan |q| \ran} =1 -T/T_{c}
\ee
so that $F$ is exactly constant.

\begin{figure}[htbp]
\begin{center}
\leavevmode
\epsfysize=250pt
\epsffile{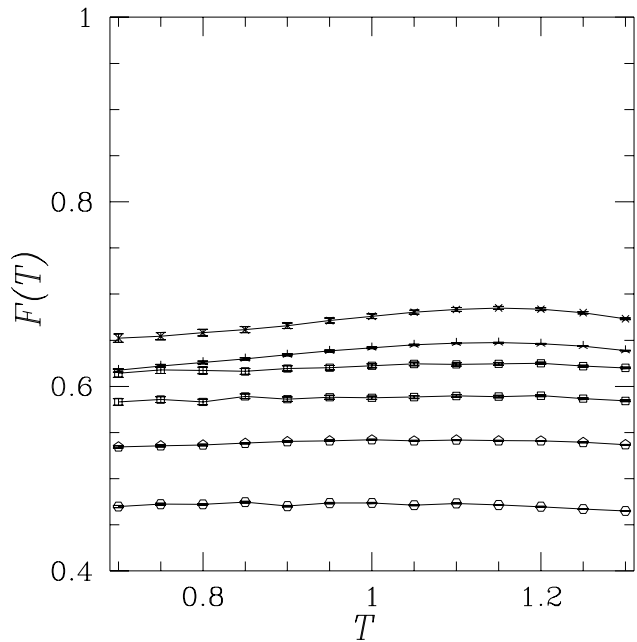}
\end{center}
\caption[0]{\protect\label{fig:sum_rule}
$F(T)$ versus $T$ for $L=4\ldots 16$.}
\end{figure}

This behavior will be in any cases changed by renormalization at
finite $D$; $\overline{\lan |q| \ran}$ does not go to zero but it is
proportional to $(1-T/T_{c})^{\beta}$, ,where the the exponent $\beta$
is equal to $\frac12 \nu (1-\eta)$, which in three dimensions is not
far from $1$ (our best estimate is $\beta= 1.30\pm 0.12$).

It is interesting to study how much in $3D$ this function stays close
to a constant in the cold phase also because it should equal to the
magnetic susceptibility.  We plot $F$ in figure (\ref{fig:sum_rule})
for different values of $L$: the function turns out to be remarkably
constant for $T<T_{c}$, where $q$ has a highly non-trivial behavior.

\section{\protect\label{S_COR}The Correlation Functions}

In reference \cite{MAPARURI} the $q=0$ ergodic component of the 
overlap correlation function was computed by using an off-equilibrium 
dynamics.  The off-equilibrium approach was used since it is difficult 
to thermalize large lattices in the cold phase, and an extrapolation 
procedure was used to deduce the equilibrium decay rate on the 
relevant distance scales.  The power law decay of this correlation 
function gives strong support to the validity of the mean field 
picture in finite dimensional systems.  Here we are able to compute 
the same correlation functions directly at equilibrium, and to show 
that we find a completely consistent result, giving in this way even 
larger support to the validity of the RSB Ansatz for describing 
realistic spin glasses.

We recall the definition of the equilibrium non-connected correlation
 function.  For each pairs of configuration $\sigma$ and $\tau$ we
 define the correlation $C(x)$ as

\be
C(x)=\frac{1}{L^3} \sum_{y}\si(x+y)\si(y)\tau(x+y)\tau(y) \ .
\ee
It may be convenient to recall that $C(x)$ does not change under a
global flip of the $\si$ or $\tau$ configurations. If we restrict the
statistic over configurations that have a given overlap $q$, the
correlation function that we will 
obtain will be denoted by $C_q(x)$ or $C(x|q)$.

\begin{figure}[htbp]
\begin{center}
\leavevmode
\epsfysize=250pt
\epsffile{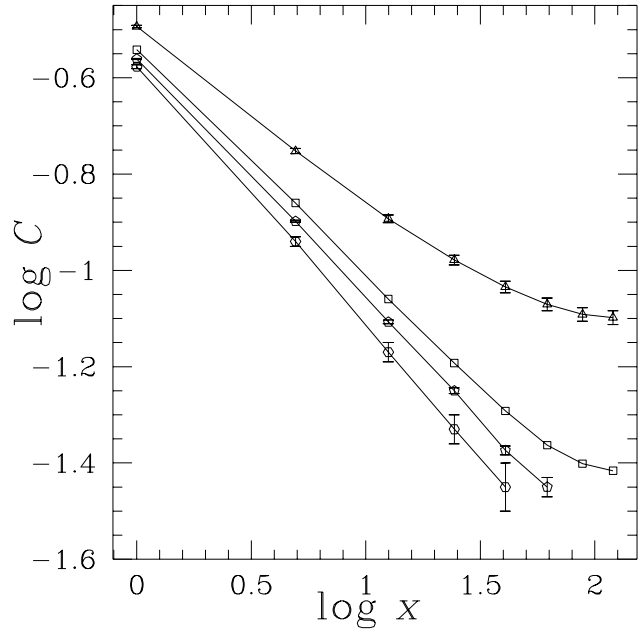}
\end{center}
\caption[0]{
  \protect\label{fig:cor_asin}
  The lower curve is the infinite time extrapolation of the 
  non-equilibrium correlation function $C_d(x|q=0)$ obtained by a 
  sudden quench ($L$ = $64$).  The second curve from the bottom is 
  $C_d(x|q=0)$ obtained by a slow annealing ($L$ = $64$).  The third 
  curve is the equilibrium correlation function computed with the 
  constraint $|q|<0.01$ ($L$ = $16$).  The upper curve is the full 
  equilibrium correlation function, including all configurations ($L$ 
  = $16$).
}
\end{figure}

The upper curve in figure (\ref{fig:cor_asin}) is the equilibrium
correlation function at $T=0.7$ on the $L=16$ lattice.  The two lower
curves are off-equilibrium $q=0$ correlation functions from
\cite{MAPARURI}: in the lower curve the system was suddenly quenched
from $T=\infty$ to $T=0.7$, while in the second one a slow annealing
procedure was used to bring $T$ down to $0.7$.  In both cases the
lattice size was $64$.  The third curve from the bottom has been
computed, by the same configurations used in the upper curve
(equilibrium runs at $T=0.7$ and $L=16$), including only couples of
configurations where $q<0.01$.  This is the real equilibrium $q\simeq
0$ correlation function, where no extrapolations were needed.  The
agreement with the result of the off-equilibrium runs is very good:
only at $x\simeq 8$, close to the center of the periodic lattice, we
see an expected discrepancy (the lattice of the off-equilibrium runs
had a linear size of $64$, and only close to $L=32$ boundary
conditions became relevant).  All the three curves are well compatible
with a power law decay of the form $x^{-\lambda}$ with $\lambda
\approx 0.5$ in agreement with the predictions of \cite{DEKOTE}.

As we have seen the propagator at fixed $q$ is equal to $q^{2}$ plus a
a term which goes to zero at infinity (see eq. (\ref{correlations})).  
Therefore at large distances
$x$

\be
P_{C}(C(x))={P(C(x)^{2})\over (2C(x))^{\frac12}},
\ee
producing a double peak structure, one peak corresponds to the delta
function at $q_{\rm EA}$ and the other peak comes from the singular
Jacobian (of the transformation $C\propto q^2$) at zero overlap.

If we naively assume that $C_{q}(x)= a(x)+b(x)q^{2}$ both peaks are
present also for finite $x$.  If we consider a more complicated
dependence of $C_{q}(x)$ at finite $x$ the leftmost peak may disappear
and one could be left with a non divergent structure.  A divergent
peak should however be present for large $x$.
 
In order to justify the equality among the dynamical result and the
static results at $q=0$ we can intuitively argue as follows. We define
an ``effective'' potential~\cite{F3} 
given by $-\log P(C(x))$.
In an off-equilibrium simulation with random initial conditions
all the components of the propagator start at zero. Using the
effective potential $-\log P(C(x))$ it is clear that the initial
$C(x)= 0$ values drifts to the closest minimum, that  is given by
$C_{\rm min}(x)$, i.e

\be
  \lim_{t\to \infty} C_d(x,t|q=0)=C_{\rm min} (x) \ ,
\ee
where $ C_d(x,t|q=0)$ is the dynamic correlation function computed at 
time $t$ and distance $x$ using a sudden quench from infinite 
temperature down to $T<T_c$ (so that the initial overlap is zero, fact 
that we denote with $q=0$ in the formula).

\begin{figure}[htbp]
\begin{center}
\leavevmode
\epsfysize=250pt
\epsffile{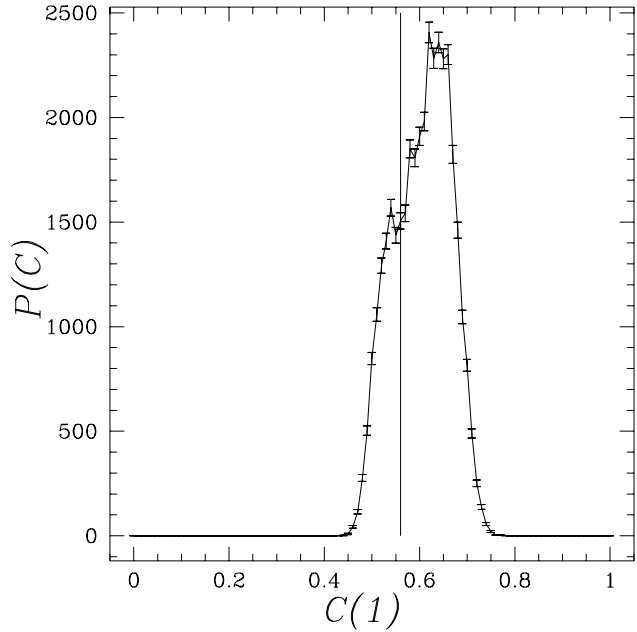}
\end{center}
\caption[0]{\protect\label{fig:hist1}
Histogram of the equilibrium correlation function at distance
$x=1$. We have marked with a vertical line the value of $C_d(1,t|q=0)$
extrapolated to infinite time \cite{MAPARURI}.}
\end{figure}

\begin{figure}[htbp]
\begin{center}
\leavevmode
\epsfysize=250pt
\epsffile{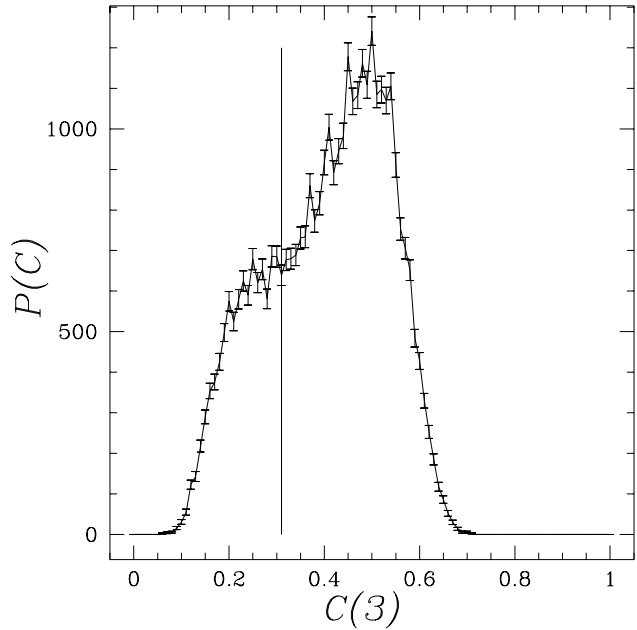}
\end{center}
\caption[0]{\protect\label{fig:hist3}
Histogram of the equilibrium correlation function at distance
$x=3$. We have marked with a vertical line the value of $C_d(3,t|q=0)$
extrapolated to infinite time \cite{MAPARURI}.}
\end{figure}

 The equilibrium probability distribution of $C(x)$ should show a
double peak structure.  The smaller peak should be located at
$C_{\min}(x)$, and this value must be equal to the off-equilibrium
value $C_d(x,\infty|q=0)$ in the infinite time limit (i.e.  the lower curve
of figure (\ref{fig:cor_asin})).  We show these histograms in figures
(\ref{fig:hist1}) and (\ref{fig:hist3}) for $x=1$ and $x=3$
respectively.  We have also marked with a vertical line the value
obtained from the off-equilibrium simulations.

Qualitatively we see that both histograms have a clear peak (for large
values of $C$, that correspond to $C_{\rm max}$) and a second maximum
or flex point (for lower values of $C$, that correspond with $C_{\rm
min}$).  We have also shown the value of the infinite time
extrapolation of $C_d(x,t|q=0)$ (the vertical line in the plots).  The
same pattern holds up to $x=5$ (that is the larger distance that we
analyzed in the dynamical runs of \cite{MAPARURI}).  It is clear from
figures (\ref{fig:hist1}) and (\ref{fig:hist3}) that the value of
$C(x|q=0)$ is located close to the second maximum or flex point.

This fact strongly supports the correctness of the off-equilibrium
approach for the computation of the $q=0$ component of the propagator
(including the effectiveness of the extrapolation procedure) and
confirms that it is possible to compute equilibrium expectation values
in off-equilibrium simulations.  Moreover the double peak structure of
the correlation function probability distribution survives in the
infinite volume limit since the dynamical expectation values have been
computed on a very large lattice (infinite to all practical effects).
It is remarkable that the distribution probability of $C(x)$ becomes
wider with increasing $x$.

All these features of the correlation functions provide a further
evidence of the existence of a non trivial probability distribution of
the overlaps, and of the fact that finite size effects are well under
control.

Finally we can check our previous estimate of the $\eta$ exponent
($\gamma/\nu=2-\eta$, and so $\eta=-0.36(6)$). We recall that at the
critical point we have

\be
C(x) \propto \frac{1}{x^{D-2+\eta}} \ .
\ee
This formula always holds at the critical point.

\begin{figure}[htbp]
\begin{center}
\leavevmode
\epsfysize=250pt
\epsffile{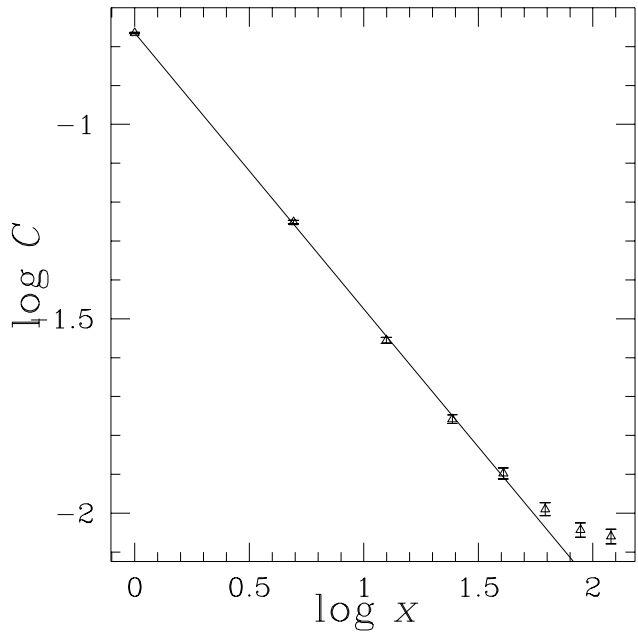}
\end{center}
\caption[0]{\protect\label{fig:cor_tc}
The correlation function near the critical point ($T=1.0$) for a $L=16$
lattice in a double (natural) logarithmic scale. We have marked with a straight
line the results of a power law fit (taking the points in the interval
$[1,5]$). The slope is in good agreement with the results found
by fitting the susceptibility as a function of the lattice size. See
the text for more details.}
\end{figure}

We show in figure (\ref{fig:cor_tc}) the overlap-overlap correlation
function near the critical point ($T=1.0$) for $L=16$. Taking
into account the points in the interval $[1,5]$ we have found a very
good power law fit with exponent $1+\eta=0.710(5)$, giving $\eta\simeq -0.29$
according with our previous estimate. Obviously for $x>5$ the
propagator is seeing the periodic boundary conditions and the
propagator misses the pure power law behavior. We can remark that this
early power law behavior that we have found (for $x \ge 1$) have been
seen, for instance, in the four and six
dimensional Ising spin glass \cite{4DIM,MEANFIELD}.

\section{\protect\label{S_CONCLU}Conclusions}

The numerical simulations we have discussed here have allowed to
establish some important results, and to make clear some relevant
issues.  Maybe the main point is that the observed equilibrium
behavior of the $3D$ spin glass shares the crucial features of RSB
solution of the mean field theory.  We are able to establish that on
large lattices, deep in the broken phase.

The study of the overlap Binder cumulant and of the spin glass
susceptibility allow a precise determination of the critical
exponents.  The analysis of sample to sample variations $P(q)$ allows
to exhibit clear power law behaviors in the probability distributions
that are typical of the mean field theory (and would have no reason to
appear in any theory of an usual ferromagnet).  Correlation functions
are now under control even at equilibrium on large lattices, for the
whole set of stable configurations and for the zero overlap part of
the phase space.

We believe that a very clear picture is emerging.

\section*{\protect\label{S_ACKNOWLEDGES}Acknowledgments}

J. J. Ruiz-Lorenzo is supported by an EC HMC (ERBFMBICT950429) grant.
We thank I. Campbell for interesting discussions.


\begin{thebibliography}{99}
\bibitem{F1} For $g_{1}=.68$
fixing $\nu=2.15$ we obtain $T_c=0.93(4)$, 
while fixing $\nu=1.85$ we find $T_c=0.95(4)$.
\bibitem{F2} It is 
  easy to obtain this formula. In terms of the weights of the 
  pure states $w_\alpha^J$ the function $P_J(q)$  is given by
        
  \be
    P_J(q)=\sum_{\alpha,\beta} w_\alpha^J w_\beta^J
    \delta(q-q_J^{\alpha,\beta})\ ,
  \ee
  where $\alpha$ and $\beta$ run over all the pure states. Now,
        
  \be
    y_J(q_c)= \int_{q_c}^1 \d q' P_J(q')= 
    \sum_{\alpha. \beta}  w_\alpha^J w_\beta^J \theta(q^{\alpha,
    \beta}_J -q_c)=\sum_I \sum_{\alpha(I), \beta(I)} w_{\alpha(I)}^J
     w_{\beta(I)}^J
     \  , 
     \protect\label{weight_y_foot}
  \ee
  where the index $I$ runs over the clusters (defined with the reference 
  value $q_c$) and $\alpha(I)$ and $\beta(I)$ run over all the pure 
  states belonging to the cluster $I$.  Taking into account the fact that 
  $w_I^J=\sum_{\alpha(I)} w_{\alpha(I)}^J$ we finally obtain
        
  \be
    y_J(q_c)= \sum_I (w_I^J)^2 \ .
  \ee
\bibitem{F3} The partition function is

  \be
    {\cal Z} = \int \d[\phi]\  \exp\left[-{\cal H}[\phi]\right] \ .
    \protect\label{E-FOOTZ}
  \ee
  For a given distance $x$ the value of the correlation
  function at distance $x$ is a function of the field $\phi$:
  $C_x=G(\phi)=V^{-1}\int dy \phi(x+y) \phi(y)$, $V=\int dy$ 
  is the volume of the system.  
  Introducing the factor $1=\int \d C_x\ \delta[C_x-G(\phi)]$ 
  in (\ref{E-FOOTZ}) we obtain

  \be
    {\cal Z} = \int \d C_x \int \d[\phi]\ 
    \delta[C_x-G[\phi]] \exp\left[ -{\cal
    H}[\phi]\right] \ .
  \ee
  We define 

  \be
    P[C_x] \equiv \int \d[\phi]\  
    \delta[C_x-G[\phi]] \exp\left[-{\cal H}[\phi]\right]
  \ee
  the probability of finding a given value of $C_x$. This allows
  to define the effective free energy (or effective
  potential) $F[C_x]$

  \be
    F[C_x] \equiv -\log P[C_x] \ ,
  \ee
  such that

  \be
    {\cal Z} = \int \d C_x \exp\left[ -F[C_x] \right]\ .
  \ee
  In the thermodynamical limit the partition function will be
  dominated by the minima of $F[C_x]$, that 
  justifies calling it an ``effective potential''.
\end{thebibliography}
\end{document}